\documentclass[a4paper]{ESASPCS13Style}
\usepackage{epsfig}

\begin{document}

\title{Multiwavelength study of X-ray emitting A- and B-stars \newline testing the companion hypothesis}

\author{B. Stelzer\inst{1} \and N. Hu\'elamo\inst{2} \and S. Hubrig\inst{2} \and G. Micela\inst{1} \and H. Zinnecker\inst{3} \and E. Guenther\inst{4}} \institute{INAF - Osservatorio Astronomico di Palermo, Piazza del Parlamento 1, I-90134 Palermo, Italy \and European Southern Observatory, Casilla 19001, Santiago 19, Chile \and Astrophysikalisches Institut Potsdam, An der Sternwarte 16, D-14482 Potsdam, Germany \and Th\"uringer Landessternwarte Tautenburg, Sternwarte 5, D-07778 Tautenburg, Germany}

\maketitle 

\begin{abstract}
No mechanism is known that produces X-rays in late B-type and early A-type
  stars.
  Nevertheless their detection has been reported from virtually all X-ray
  satellites, and has remained a mystery to date.
  We use a multi-wavelength approach to test the most widespread hypothesis
  that the X-rays are generated by late-type magnetically active companions.
  Our high spatial resolution observations of A/B-type stars in the IR using
  adaptive optics uncover hypothetical companion stars at arcsecond
  separations from the primary. The same targets are then followed-up in X-rays with
  {\em Chandra}. {\em Chandra}'s unprecedented spatial resolution allows to check whether
  the new infrared sources are responsible for the X-ray emission
  previously ascribed to the A/B-type star. Finally, those A/B-type stars
  that are still detected with Chandra are studied with IR spectroscopy,
  where we search for temperature sensitive features indicating the existence
  of even closer cool companions.
  Even with this multi-fold strategy we are likely to miss the closest of the
  possible companions, but a study of the X-ray properties can provide
  further information on the nature of the emitters.
\end{abstract}

\section{Introduction}
  
For stars on the main-sequence two mechanisms are known to generate X-ray emission:
In O- and early B-type stars the X-rays are produced
by instabilities arising in the strong radiatively driven stellar
winds (\cite{Owocki99.1}, \cite{Lucy80.1}), and in late-type stars a
solar-like magnetic dynamo is thought to produce the observed X-ray
activity (\cite{Parker55.1}, \cite{Parker93.1}, \cite{Ruediger95.1}). 
No X-ray emission is expected from stars whose spectral types are late B and 
early A, because they do not drive strong enough winds nor do they possess
convective zones necessary to sustain a magnetic dynamo. Therefore, 
their X-ray detection which has repeatedly been reported throughout the 
literature 
(e.g. \cite{Grillo92.1}, \cite{Schmitt93.2}, 
\cite{Simon95.1}, \cite{Berghoefer96.1},
\cite{Panzera99.1}, \cite{Huelamo00.1}, 
\cite{Daniel02.1}) 
has remained a mystery to date. 

Lacking any theoretical model for X-ray production intrinsic to intermediate-mass 
stars, the observed emission is commonly attributed to unresolved late-type companions.
In order to check this hypothesis \cite*{Berghoefer94.1} carried out 
{\em ROSAT} High Resolution Imager (HRI) X-ray observations   
of visual binaries composed of late-B type stars and known visual companions at 
separations $>\,10^{\prime\prime}$, i.e. those
clearly resolvable by the HRI. In these observations only in $1$ out of $8$ 
cases the X-ray emission could be ascribed to the late-type companion. On the other
hand, a {\em ROSAT} HRI study of visual binary systems comprised
of early-type stars and post-T Tauri stars (also known as Lindroos
systems), has shown that both the late-B type primaries and their
late-type companions emit X-rays at similar levels 
(\cite{Schmitt93.2}, \cite{Huelamo00.1}).  The similarity of the X-ray
properties of the Lindroos primaries and secondaries supports the
hypothesis that the X-ray emission from the late B-type stars in fact 
originates from closer late-type companions unresolvable by the
{\em ROSAT} HRI. Since a large fraction of the X-ray detected late B-type stars
belong to rather young ($\sim 10^{7...8}$\,yrs) 
stellar groups (e.g., Sco\,OB2, Carina-Vela, Tucanae),
most of the predicted unresolved late-type stars may be young stars still
contracting to the main-sequence (MS) or just arrived on the zero-age MS, 
if bound to the primaries. 

Thanks to an observationally established but theoretically poorly understood 
enhancement of magnetic activity at young stellar ages, 
late-type pre-MS stars ( = T Tauri stars) are ubiquitous X-ray sources. 
The picture is less clear for their higher-mass counterparts, the HAeBe stars. 
When starting their evolution fully convective they may drive magnetic activity
by the same process as T Tauri stars, or by magnetic interaction
between the star and an accretion disk (as proposed by \cite{Montmerle00.1} for proto-stars). 
\cite*{Tout95.1} argue that intermediate-mass stars may be able to maintain dynamo
action also throughout the first part (several percent) of their radiative phase. 
On the other hand, close visual companions to HAeBe stars have been presented by \cite*{Li94.1},
\cite*{Pirzkal97.1}, and \cite*{Leinert97.1}. Monte Carlo simulations by \cite*{Pirzkal97.1}
have suggested that almost all HAeBe stars have companions within a completeness limit of
$0.4^{\prime\prime}<$ sep $<8^{\prime\prime}$ and $K < 10.5$. 
The X-ray emission from HAeBe stars has been investigated systematically by 
\cite*{Damiani94.1} based on {\em EINSTEIN} observations and by \cite*{Zinnecker94.1} 
using {\em ROSAT}. About $30$\,\% and $50$\,\% of the observed HAeBe stars were detected, respectively.
According to \cite*{Zinnecker94.1} the most plausible origin of these detections are their winds. 
On the other hand, X-ray flares have been detected on a small number of HAeBe stars
(\cite{Hamaguchi00.1}, \cite{Giardino04.1}), and have been taken as 
evidence for magnetic activity. {\em EINSTEIN} and {\em ROSAT} observations did not resolve
the HAeBe stars from the companions identified in the publications cited above.

\section{Observing Strategy}

Solving the puzzle of X-ray emission from intermediate-mass stars calls for a
complex observational approach involving imaging and spectroscopic
observations in different wavelength bands. Our strategy to test the companion
hypothesis is as follows:
(i) search for cool companions to A- and B-type stars with high-resolution 
imaging observations in the IR using the adaptive optics (AO) technique, (ii) follow-up
X-ray observations with similarly high spatial resolution to pinpoint the X-ray 
emitter in the newly identified systems, (iii) IR spectroscopy for those
intermediate-mass stars that are X-ray detected even after being resolved from 
any (sub-)arcsecond visual companions 
to search for even closer companions. 

Throughout this article, we will for simplicity call any faint IR object near 
the intermediate-mass stars of interest `companions'. It should be kept in mind, however, 
that the objects newly discovered in AO imaging have not been confirmed yet to be physically bound to 
the `primaries'. True late-type companions to MS B-type stars are expected to be young, because of
the different evolutionary time-scales of early- and late-type stars. Optical spectroscopy should 
reveal a Li\,I absorption feature at $6708$\,\AA~ in physical companions proving their pre-MS nature. 
But no observations have been carried out yet to that effect.

Work on the three sub-projects introduced above has started in parallel. Here we report on the
first results of  
\begin{itemize}
\item AO imaging of early-A type stars in the northern hemisphere at the 
Telescopio Nazionale Galilei (TNG) on La Palma;
\item {\em Chandra} X-ray observations of late-B type stars for which the recent AO studies of
\cite*{Hubrig01.1} and \cite*{Huelamo01.1} have identified AO companions; 
\item IR spectroscopy of X-ray emitting late-B type stars at the Th\"uringer Landessternwarte 
Tautenburg. 
\end{itemize}

%
%
%
\begin{figure*}[ht]
\begin{center}
\parbox{16cm}{
\parbox{8cm}{\epsfig{file=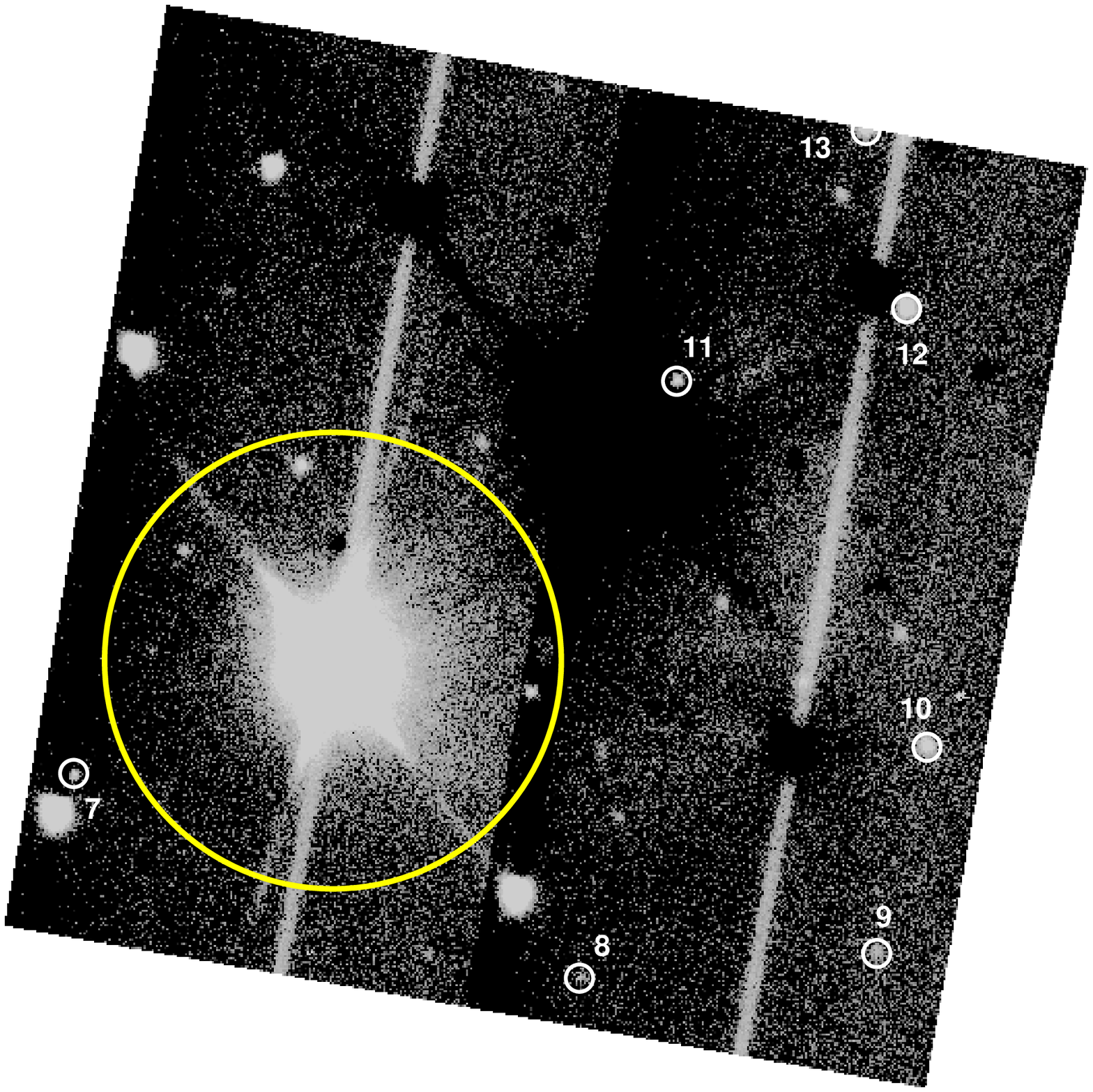, width=8cm}}
\parbox{8cm}{\epsfig{file=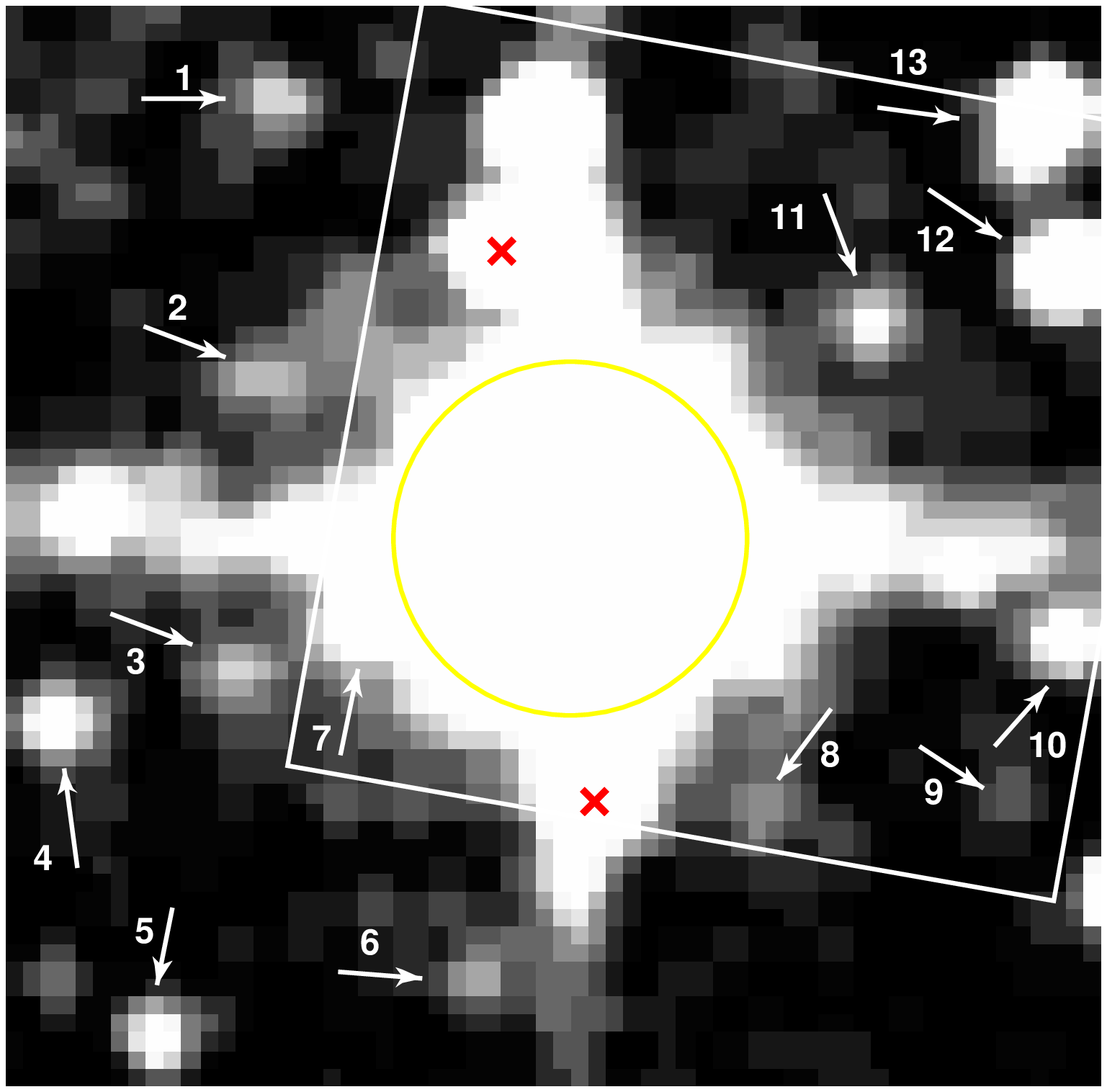, width=8cm}}
}
  \end{center}
\caption{$K$-band images of HR\,7826. {\em left} - TNG/NICS+AdOpt, $0.7^\prime \times 0.7^\prime$; 
the black features are negatives resulting from the subtraction of two science frames, 
as well as the three white features forming a square with the main negative; the faint white objects 
are real, and those of them labeled with numbers can be identified on the 2\,MASS image to the right. 
{\em right} - 2\,MASS, $1^\prime \times 1^\prime$; the area of the TNG image is marked by the slightly 
tilted squared frame; dark crosses are artifacts, arrows with numbers are 2\,MASS sources.}\label{fig:tng}
\end{figure*}

\section{Results}\label{sec:results}

\subsection{IR Imaging with Adaptive Optics}\label{subsec:ao}

Previous AO searches for faint companions to (X-ray emitting) intermediate-mass MS stars have been 
carried out in the southern hemisphere 
(see \cite{Hubrig01.1}, \cite{Huelamo01.1}, and \cite{Shatsky02.1}).
These studies have provided a wealth of new candidate companions, 
but have been restricted to the higher-mass end of the `X-ray forbidden' range of spectral
types. Since the final aim is to prove/disprove the existence of the theoretically predicted gap in 
X-ray emission along the spectral type sequence, and to understand its point of onset, we must sample 
the whole critical range of spectral types.  
Therefore we have engaged in AO observations of A-type stars. The sample
was selected from the Catalogue of Optically Bright Main-Sequence Stars detected
during the {\em ROSAT} All-Sky Survey (\cite{Huensch98.1}). 
We have chosen stars with spectral types ranging between A0 and A5 and
without indications of binarity according to the {\em Hipparcos} and Wielen
Catalogues (\cite{Turon93.1}, \cite{Wielen00.1}).  With this latter
criterion we minimize the probability that spectroscopic binaries are
included in our sample. In contrast to previous studies our targets are located
in the northern sky. 

So far we have obtained $H$- and $K$-band images of $23$ early-A type stars 
using the small field of the Near Infrared Camera Spectrometer (NICS) at the TNG.
Combined with the AO system the field-of-view is $0.7^\prime \times 0.7^\prime$. 
As a first result we show in Fig.~\ref{fig:tng} 
the reduced frame for HR\,7826, an A3\,V star with no reports on binarity in the literature. 
Dark patches result from the subtraction of the science frames effected to eliminate the
contribution of the sky background. The white (black) objects that -- together with the main negative  
(positive) -- delineate a square-shaped region are ghosts. Potential companion stars are to be 
found among the remaining bright point-like features. Indeed, several of them have already
known 2\,MASS counterparts, marked with white circles and numbers for identification with the
2\,MASS image aside. The large circle overplotted on
the image denotes a separation of $10^{\prime\prime}$ from the B-type star. 
On the right hand side in the same figure we show the corresponding 2\,MASS image. Obviously, 
in 2\,MASS no faint sources can be detected within the critical region of $\sim 10^{\prime\prime}$ 
around the bright B-star due to the poor spatial resolution of this survey. 

A thorough analysis of this TNG image and those of the other targets will produce a substantial 
list of new companion candidates. These new objects shall then be considered in follow-up 
{\em Chandra} observations and optical spectroscopy.

\subsection{High Spatial Resolution X-ray Follow-up}\label{subsec:xray}

\begin{figure*}[ht]
  \begin{center}
\parbox{16cm}{
\parbox{8cm}{\epsfig{file=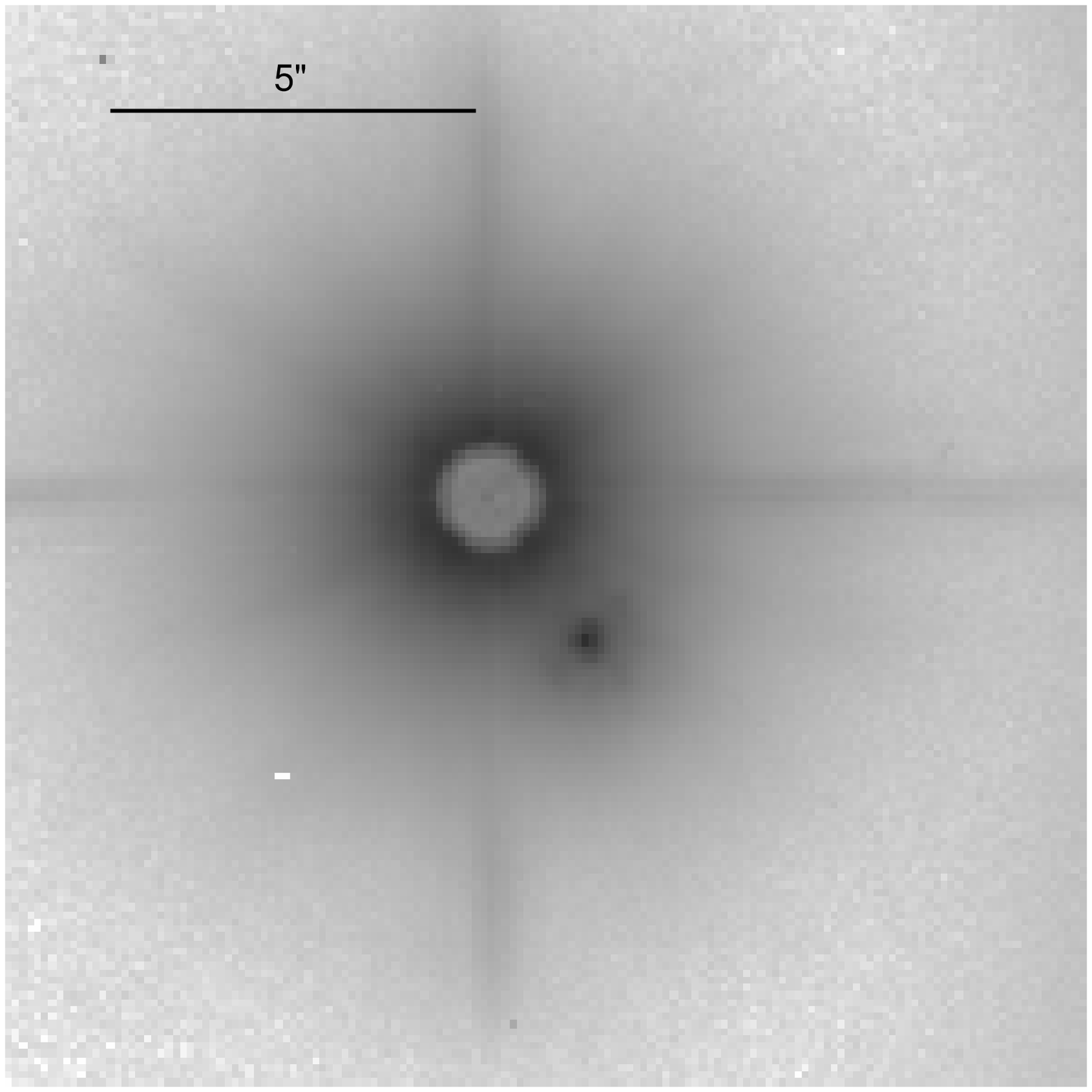, width=7cm}}
\parbox{8cm}{\epsfig{file=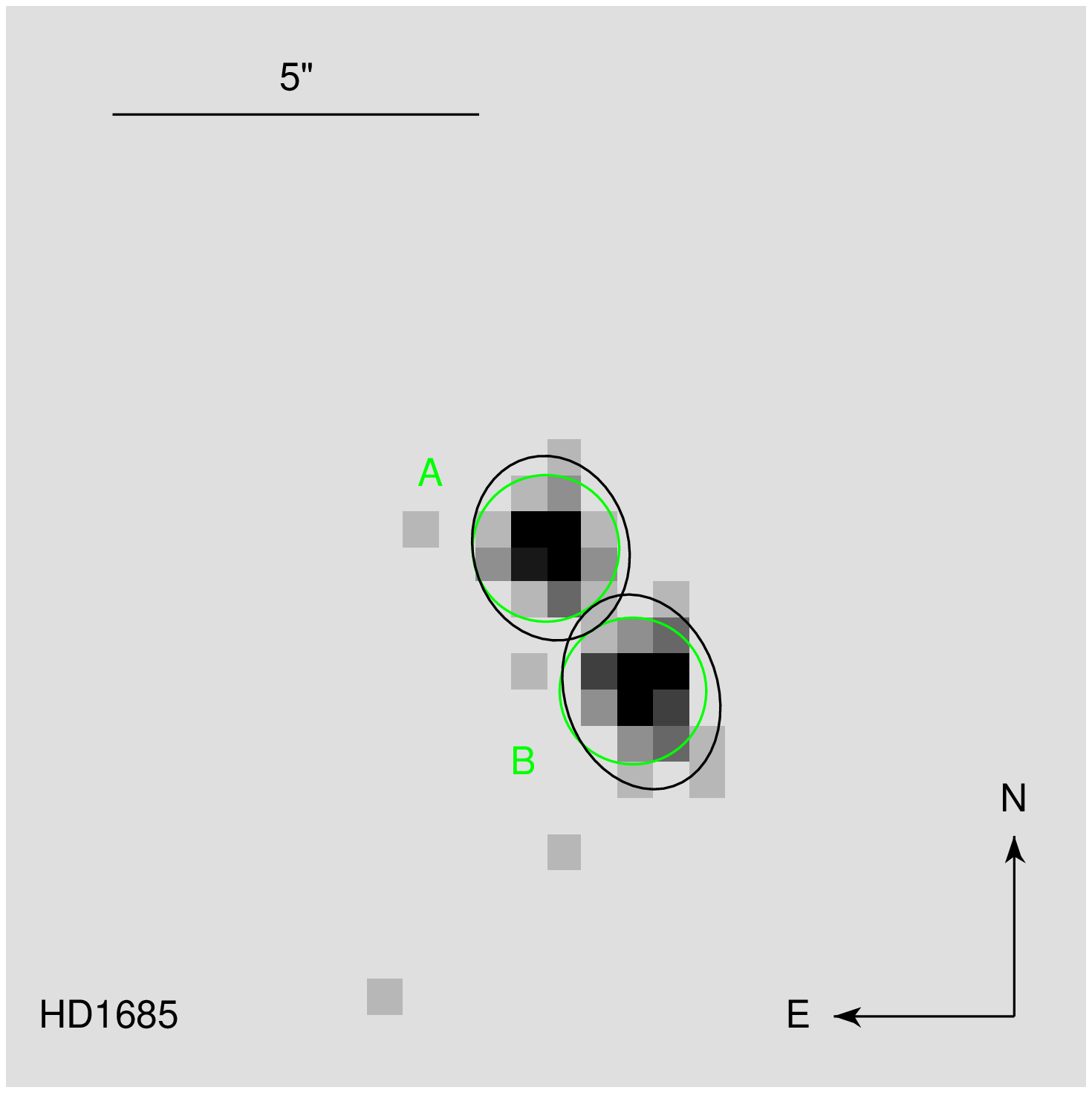, width=8cm}}
}
  \end{center}
\caption{Images of HD\,1685: {\em Left} - ADONIS ESO/3.6m, $K$-band discovery image of a faint IR object $2.3^{\prime\prime}$ south-west of the B-type primary; {\em right} - {\em Chandra} ACIS image showing that both objects are detected, `A' is the primary, `B' is the new IR object.}\label{fig:ao_xray_images}
\end{figure*}

{\em Chandra} is the first and so far only X-ray satellite providing spatial resolution 
comparable with AO observations. In addition it provides spectral
capabilities that give information about the properties of the X-ray emitting plasma,
such as temperature and luminosity, which can be used to constrain the nature of the target. 

Within this project $9$ late-B type stars on the MS 
were observed with {\em Chandra}'s Advanced CCD Imaging Spectrometer (ACIS). 
The targets have shown to be X-ray emitters in the {\em ROSAT} All-Sky Survey (= RASS; \cite{Berghoefer96.1}) 
and they have AO companions from the work of \cite*{Hubrig01.1} with separations ranging from 
$\sim 1-8^{\prime\prime}$, i.e. the systems are well resolvable with {\em Chandra}. 
This sample is complemented by data of two stars extracted from the {\em Chandra} archive,
that obey the same criteria: spectral type late-B, on the MS, X-ray sources according to the
RASS, and close companions resolvable with {\em Chandra}.
Some of the targets are Lindroos systems (\cite{Lindroos85.1}), and have additional companions at wider 
separations. We searched for X-rays also at the position of these Lindroos secondaries.

In addition we scanned the {\em Chandra} archive for any observations of HAeBe stars, and
found $15$ of them. Some of these data have been published by \cite*{Feigelson03.1} and \cite*{Giardino04.1}, 
others are serendipitous sources of pointings with a different scope.
We point out that the HAeBe star sample diverges from the original selection criterion, in that 
many of them are not known to have close visual companions. 
But they are of interest to our study because the
physical processes related with their X-ray emission may be different from those of the more evolved MS 
B/A-type stars. A comparison between X-ray properties of HAeBe and MS B/A-type stars should allow to test 
whether and where there is an age-related transition or shut-off in the dynamo action.

An example for the {\em Chandra} imaging observations is given in Fig.~\ref{fig:ao_xray_images}, which shows 
the ACIS exposure of HD\,1685, together with the corresponding AO discovery image of the `companion' 
at the 3.6\,m telescope of ESO (Chile). 
Two X-ray sources are seen, that can be identified with the B-type primary and the AO companion,
respectively. 

To summarize, almost all of the new IR sources turn out to be X-ray emitters:  
only one of them is undetected with {\em Chandra}, but it is probably a faint source just below 
the detection sensitivity (see discussion in \cite{Stelzer03.1} where the first $5$ targets have been
presented). However, this does not prove
the companion hypothesis, as we have detected also $7$ out of $11$ B-type stars. 
One of them is known to be a spectroscopic binary, the other $6$ B-type stars remain candidates for
being intrinsic emitters until tested for closer companions with IR spectroscopy 
(see Sect.~\ref{subsec:irspec}). 

Previously known Lindroos companions at separations $\geq 10^{\prime\prime}$ are present in 
HD\,113703, HD\,129791, HD\,32964 and HD\,123445. The latter two are undetected, suggesting that they
may not form bound systems with the `primary'. Indeed -- consistent with our X-ray observations -- 
the Lindroos companions of HD\,32964 and HD\,123445
have been labeled as likely optical pairs, while the Lindroos companions of HD\,113703 and HD\,129791
have been dubbed likely physical based on their optical photometry and spectroscopy 
(\cite{Eggen63.1}, \cite{Pallavicini92.1}).

Among the HAeBe stars $12$ out of $15$ are detected with {\em Chandra}. Whether the higher detection fraction
with respect to the MS B-type stars indicates a different emission mechanism or is the result of
hidden companions remains unclear so far. If all X-rays from the position of intermediate-mass stars 
are assumed to be generated by late-type companion stars 
the higher detection rate of HAeBe stars could also be due to higher activity levels of their companions,
because of their younger age with respect to the companions of MS stars. 

Table~\ref{tab:data} gives a summary of the sample of {\em Chandra} targets, including information on
the components in the multiples and two flags estimating the companion status based on X-ray data and
near-IR photometry. The upper part of the table lists MS stars, and the lower part HAeBe stars. 
\begin{table*}[bht]
\caption{Intermediate-mass stars observed with {\em Chandra} and known companions or companion candidates
at arcsecond separations (Lindross companions are labeled `L'; separation and position angle are given in columns~4 and~5). 
The last two columns provide flags for the companion status: 
`$\surd$' - X-ray detection or near-IR photometry suggests late-type pre-MS star and consequently likely bound system, 
`?' - no near-IR color available, `N.A.' - primary, the flag is not applicable.}
  \label{tab:data}
  \begin{center}
    \leavevmode
    \footnotesize
    \begin{tabular}[h]{llccrcc}\hline \\[-5pt]
      Designation & SpT      & Compon. & Sep                & P.A.       & \multicolumn{2}{c}{Companionship} \\[+5pt]
                  &          &       & [${\prime\prime}$] & [$^\circ$] & X-rays & NIR phot. \\[+5pt]\hline \\[-5pt]
      \multicolumn{7}{c}{Main-sequence B-type stars and companion candidates} \\[+5pt]\hline \\[-5pt]
      HD\,1685    & B9   & A &      &       &  N.A.    & N.A. \\
      HD\,1685    & $-$  & B & 2.28 & 211.4 &  $\surd$ & $-$ \\
      HD\,113703  & B5   & A &      &       &  N.A.    & N.A. \\
      HD\,113703  & $-$  & B & 1.55 & 268.2 &  $\surd$ & $\surd$ \\
      HD\,113703  & K0   & L & 11.5 &  79   &  $\surd$ & ? \\
      HD\,123445  & B9   & A &      &       &  N.A.    & N.A. \\
      HD\,123445  & $-$  & B &5.56/5.38&65.0/64.0 & $\surd$ & $\surd$ \\
      HD\,123445  & K2   & L & 28.6 &  35   &  $-$     & $\surd$ \\
      HD\,133880  & B8   & A &      &       &  N.A.    & N.A. \\
      HD\,133880  & $-$  & B & 1.22 & 109.2 &  $\surd$ & $\surd$ \\
      HD\,169978  & B7   & A &      &       &  N.A.    & N.A. \\
      HD\,169978  & $-$  & B & 3.09 & 168.7 &  $-$     & ? \\
      HD\,32964   & B9.5 & A &      &       &  N.A.    & N.A. \\
      HD\,32964   & $-$  & B & 1.61 & 232.6 &  $\surd$ & $\surd$ \\
      HD\,32964   & K5   & L & 52.8 &  10.0 &  $-$     & $\surd$ \\     
      HD\,73952   & B8   & A &      &       &  N.A.    & N.A. \\
      HD\,73952   & $-$  & B & 1.16 & 205.3 &  $\surd$ & $\surd$ \\
      HD\,110073  & B8   & A &      &       &  N.A.    & N.A. \\
      HD\,110073  & $-$  & B & 1.19 &  75.0 &  $\surd$ & $\surd$ \\
      HD\,134837  & B8   & A &      &       &  N.A.    & N.A. \\
      HD\,134837  & $-$  & B & 4.70 & 154.3 &  $\surd$ & ? \\
      HD\,134946  & B8   & A &      &       &  N.A.    & N.A. \\
      HD\,134946  & $-$  & B & 8.21 &  45.3 &  $\surd$ & ? \\
      HD\,129791  & A0   & A &      &       &  N.A.    & N.A. \\ 
      HD\,129791  & $-$  & L & 35.3 & 205.5 &  $\surd$ & $\surd$ \\[+5pt]\hline \\[-5pt] 
      \multicolumn{7}{c}{HAeBe stars and companion candidates} \\[+5pt]\hline \\[-5pt]
      HD\,104237  & A    & A &      &       &  N.A.    & N.A. \\
      HD\,100546  & B9   & A &      &       &  N.A.    & N.A. \\
      HD\,100546  & $-$  & B & 4.54 & 196.5 &  $-$ & $-$ \\
      HD\,100546  & $-$  & C & 5.22 & 155.1 &  $-$ & $-$ \\
      HD\,100546  & $-$  & D & 5.91 &  26.4 &  $-$ & ? \\
      HD\,100546  & $-$  & E & 5.55 & 322.6 &  $-$ & ? \\
      HD\,141569  & B9   & A &      &       &  N.A.    & N.A. \\
      HD\,141569  & $-$  & B & 7.57 & 311.5 &  $\surd$ & $\surd$ \\
      HD\,141569  & $-$  & C & 8.93 & 310.0 &  $\surd$ & $\surd$ \\
      HD\,150193  & A1   & A &      &       &  N.A.    & N.A. \\
      HD\,150193  & $-$  & B & 1.10 & 236   &  $\surd$ & $\surd$ \\
      V892\,Tau   & A6   & A &      &       &  N.A.    & N.A. \\
      V892\,Tau   & $-$  & B & 4.10 &  23.4 &  $\surd$ & $-$  \\
      HD\,152404  & F5   & A &      &       &  N.A.    & N.A. \\
      HIP\,16243  & B8   & A &      &       &  N.A.    & N.A. \\
      HD\,147889  &B2III/IV&A&      &       &  N.A.    & N.A. \\
      HD\,97300   & B9   & A &      &       &  N.A.    & N.A. \\
      V380\,Ori   & B8+A1&A+B& 0.15 & 204.2 &  N.A.    & N.A. \\
      V590\,Mon   & B8   & A &      &       &  N.A.    & N.A. \\
      TY\,CrA     & B9   & A &      &       &  N.A.    & N.A. \\
      R\,CrA      & A5II & A &      &       &  N.A.    & N.A. \\
      HD\,176386  & B9IV & A &      &       &  N.A.    & N.A. \\
      MWC\,297    & O9   & A &      &       &  N.A.    & N.A. \\
      \hline \\
      \end{tabular}
  \end{center}
\end{table*}

\begin{figure}[ht]
  \begin{center}
    \epsfig{file=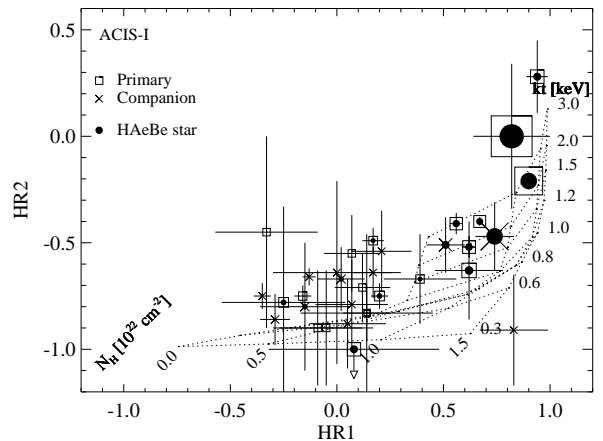, width=8cm}
  \end{center}
\caption{{\em Chandra} ACIS hardness ratios for all stars discussed in this paper with $\geq 10$\,counts in the broad band. The size of the plotting symbols is scaled to $A_{\rm V}$.}\label{fig:hr}
\end{figure}

Spectral analysis provides X-ray luminosities and an estimate of the temperature in the 
emitting region. Since the exposure times were short -- the major aim was the detection, not a detailed 
analysis -- typically only $\sim 20-100$ source photons were collected.  
In cases of such poor statistics commonly hardness ratios are used to describe spectral properties of 
X-ray sources. Hardness ratios are defined as $HR = (H-S)/(H+S)$, where $H$ and $S$ are the number of 
counts in a hard band and in a soft band, respectively.
Here, $HR\,1$ compares the $0.5-1$\,keV ($S$) and the $1-8$\,keV ($H$) band,
and $HR2$ the $1-2$\,keV ($S$) and the $2-8$\,keV ($H$) band.
In Fig.~\ref{fig:hr} we display the observed hardness ratios
super-imposed on a model grid representing a 1-T Raymond-Smith (\cite{Raymond77.1}) spectrum subject to photo-absorption.
The model grid was computed with PIMMS. The plotting symbols for the data have been scaled to the visual
extinction. The most absorbed sources (some of the HAeBe stars) are found in the right part of the diagram,
where the model indicates high column density. Thus, we conclude that the hard X-ray emission of these 
stars is not an intrinsic property, but an artifact produced by absorption of the soft component. This is
supported by the fact that the weakly absorbed HAeBe stars and the MS stars and companions can not be
distinguished in terms of their hardness ratios. The majority of the MS B-stars and their companions cluster
at intermediate values of $HR\,1$, just above the unabsorbed 1-T model. Their location outside the boundaries
of the RS-model can probably be attributed to the fact that the iso-thermal model is an
inadequate representation for their coronal temperature structure. No drastic differences in the hardness of
primaries and companions are observed, suggesting that the X-rays are produced by the same mechanism. 

\begin{figure}
\begin{center}
\epsfig{file=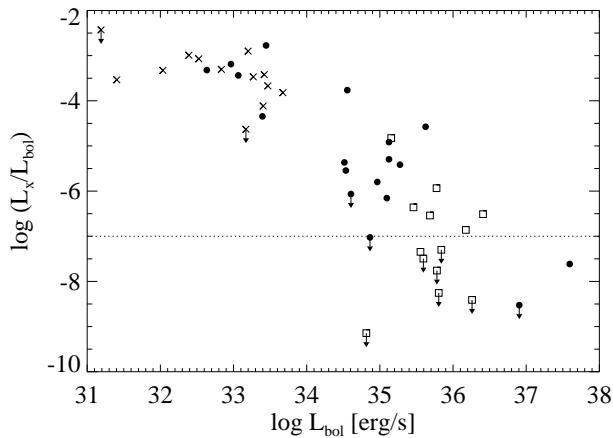, width=8cm}
\caption{Ratio between X-ray and bolometric luminosity for Chandra observed A- and B-type stars; plotting symbols as in Fig.~\ref{fig:hr}. The dotted line indicates the empirical $L_{\rm x}/L_{\rm bol}$ ratio for wind-driven stellar X-ray sources.}
\label{fig:acis_lx_lbol}
\end{center}
\end{figure}

Fig.~\ref{fig:acis_lx_lbol} displays the $L_{\rm x}/L_{\rm bol}$ ratio 
for all components of the A- and B-type stars observed so far with {\em Chandra}. 
The ratio between X-ray and bolometric luminosity is  
a crucial indicator for stellar activity. 
In general, the most active stars -- often being the most rapid rotators -- 
are observed to display values near $10^{-3}$.
An unidentified mechanism seems to prevent the generation of X-rays beyond
this limit. 
Less active late-type stars range between $L_{\rm x}/L_{\rm bol} = 10^{-4...-5}$. 
The spread is thought to be caused by the influences of various stellar parameters
such as mass, rotation, and age on the level of X-ray emission. 
O- and B-type stars, for which X-ray emission is thought to arise in a stellar wind,
are clearly distinct from late-type stars with a typical value of 
$L_{\rm x}/L_{\rm bol} \approx 10^{-7}$.

In the sample
investigated here the IR companions display $\log{(L_{\rm x}/L_{\rm bol})}$ 
values near the saturation limit.
For all intermediate-mass MS stars of the sample which are not detected the upper
limits we derive are lower than the canonical value of $10^{-7}$, making stellar winds
an unlikely cause for their production. 
The detected B-type MS stars show intermediate values of 
$\log{(L_{\rm x}/L_{\rm bol})}$ and need to be examined for the presence of further,
as yet undiscovered companions. 
The HAeBe stars fill the gap in $L_{\rm bol}$ between the intermediate-mass stars
and their low-mass companions. In terms of  $\log{(L_{\rm x}/L_{\rm bol})}$ they
also occupy an intermediate position with values of $-4...-6$, except 
for the two very luminous objects MWC\,297 and HD\,147889. 
However, we note that the bolometric luminosities of the HAeBe stars in some cases
are highly uncertain, due to excess emission from remnant circumstellar material.

\subsection{High S/N IR spectroscopy}\label{subsec:irspec}

\begin{figure}
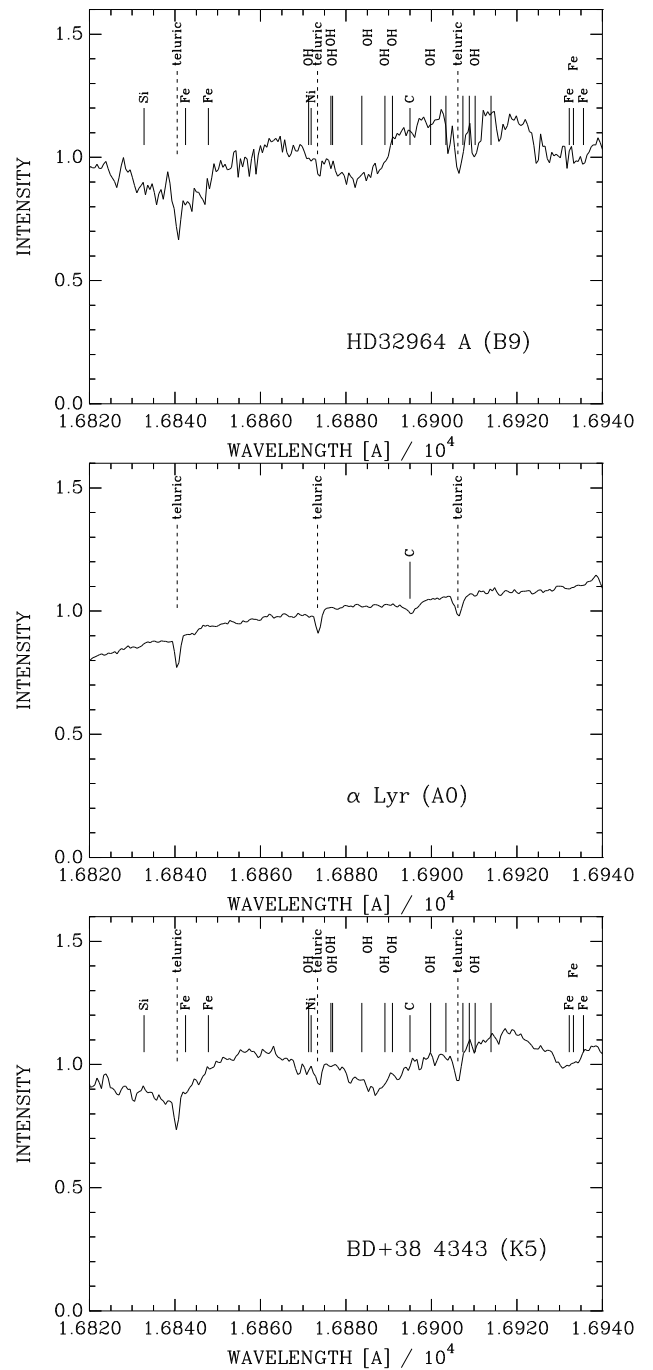

\begin{center}
\epsfig{file=./bstelzer_f5a.ps, width=6cm, angle=270}
\epsfig{file=./bstelzer_f5b.ps, width=6cm, angle=270}
\epsfig{file=./bstelzer_f5c.ps, width=6cm, angle=270}
\caption{$H$-band spectrum of science target HD\,32964, and comparison stars $\alpha$\,Lyr and BD+38-4343.}
\label{fig:ir_spec}
\end{center}
\end{figure}

The aim of this part of the project is to 
unveil signatures of late-type stars in the high S/N IR spectra of intermediate-mass stars.
The IR range is preferred over the optical because of the lower flux ratio between early-
and late-type stars, favoring the detection of the latter. 
While in a late B-type star the strongest (and only) features in the $H$- and $K$-band
spectrum are the hydrogen lines, 
in the case of a late-type star the IR spectrum shows strong absorption lines
from atomic species (Ca\,I, Mg\,I, Si\,I) and molecular bands (CO and OH). 
Therefore, this spectral range is suited to determine the spectral class of the target. 

In a pilot study we observed HD\,32964 at the 2.2-m-telescope on Calar Alto. 
HD\,32964 is a complex system composed of the following objects:
\begin{itemize}
\item the primary, a spectroscopic binary of two stars with nearly equal mass of $\sim 2.4\,M_\odot$,
\item a Lindroos companion at $53^{\prime\prime}$, spectral type K5\,V; 
\item an AO companion at $1.6^{\prime\prime}$.
\end{itemize}
The only X-ray source detected with {\em Chandra} is the AO companion. Near-IR photometry puts this object
near the zero-age MS in the color-magnitude diagram, consistent with the relatively old age ($\sim 200$\,Myr) 
derived for the primary by \cite*{Hubrig01.1}. The non-detection of the Lindroos
secondary provides support for the suggestion by \cite*{Eggen63.1} that it is probably physically 
unrelated.

For the IR spectroscopy we used the Coud\'e Spektrograph with the MAGIC infrared 
camera at the f/12 camera of the spectrograph. 
The setup provides a 2-pixel-resolution of $\lambda / \Delta \lambda = 14000$.
The wavelength region between $16820-16940$\,\AA~ was chosen, 
because it contains the OH-features which are prominent in late-type stars. 

The spectrum of HD\,32964 is shown in Fig.~\ref{fig:ir_spec} on the top. 
Next to the OH-features we mark other lines seen in the solar spectrum (\cite{Livingston91.1}) 
and thus possibly visible in the spectrum of a K-type star. 
Comparison with the spectrum of Vega (middle panel of Fig.~\ref{fig:ir_spec}), 
similar in spectral type to the primary HD\,32964\,A,
reveals major differences. The spectral shape of HD\,32964 resembles much more that of a mid-K type star 
(see comparison spectrum on the bottom of Fig.~\ref{fig:ir_spec}). 
We must caution that with a slit width of $\sim 2^{\prime\prime}$ the AO companion 
probably contributes to this spectrum. Therefore, our data does not report the discovery of an
additional spectroscopic companion, but contamination by the nearby visual companion. However, it
demonstrates that the approach works in principle and outlines that the instrumental setup must be
carefully chosen. A detailed discussion of this data set awaits further analysis. 
Observations of a subsample of southern late-B and A-type stars have been
carried out with SofI at the NTT. The results will be presented elsewhere.

\subsection{Nature of the Companions}\label{subsec:comp_nature}

\begin{figure}
\begin{center}
\epsfig{file=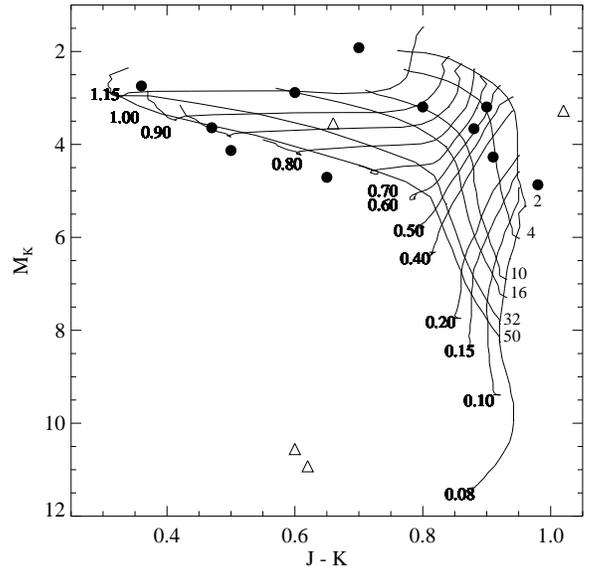, width=8cm}
\caption{Near-IR color-magnitude diagram for companion candidates to intermediate-mass stars. {\em filled symbols} - objects detected with {\em Chandra}, {\em open symbols} - objects not detected with {\em Chandra}. The grid represents pre-MS models from \protect\cite*{Baraffe98.1}. Errors are not shown for clarity, but are rather large, such that all but the two objects with the lowest $M_{\rm K}$ values in this diagram are compatible with being on the pre-MS.}
\label{fig:cmd}
\end{center}
\end{figure}

X-ray detection can help to single out true companions from chance projections. 
Of the $15$ companions to the $11$ MS B-stars observed with {\em Chandra} $12$ are detected. 
The $15$ HAeBe stars have $9$ companions in total of which $4$ are detected with {\em Chandra}. 
One object from the HAeBe sample is a close binary composed of two intermediate-mass stars. 
The remaining $4$ undetected `companions' are probably unrelated objects. However, final classification
requires confirmation of the companion status by means of spectroscopy or proper motion.  
To date this information is not available for any of the IR objects discovered with AO near the 
MS A- or B-type stars discussed in this article. In the meantime 
available near-IR photometry allows for a rough estimate of their evolutionary stage,
comparing their position in the near-IR color-magnitude diagram to pre-MS models. 

Fig.~\ref{fig:cmd} shows the $M_{\rm K}$ vs. $J-K$ diagram with model calculations by \cite*{Baraffe98.1}:
$Y=0.275$, $[M/H]=0$, $\alpha_{\rm ML}=1$. For the distances needed to compute the absolute $K$-band magnitude
we assumed that all companions are bound to the primaries. 
Note that of the presumed low-mass companions $5$ have been observed only in the $K$-band, such that they can
not be placed in Fig.~\ref{fig:cmd}. For $4$ `companions' the near-IR colors are not compatible with them
being on the pre-MS, and therefore these objects are likely physically unrelated to the A-/B-type or HAeBe star. 

We tentatively assigned a companion status based on X-ray detection/non-detection and near-IR photometry. 
Corresponding flags are given in the last two columns of Table~\ref{tab:data}.

\section{Summary and Outlook}\label{sec:summary}

We have engaged in IR and X-ray observations of inter\-mediate-mass stars
in order to examine whether late-type companion stars are the cause of their unexplained X-ray emission. 
Previous X-ray studies of known visual systems composed of B-type primary and late-type
secondary have remained inconclusive. A major part of these systems
were not resolvable with {\em ROSAT}. The exceptional spatial resolution of {\em Chandra} 
allows us now to study systems as close as $\sim 1^{\prime\prime}$. 
This enables to access in the X-ray range many of the faint IR objects discovered near B-type stars 
in recent AO surveys. Combined {\em Chandra} and AO imaging yields a large sample of homogeneous
observations on which the companion hypothesis can be tested. Our AO survey of A-type stars in the 
northern hemisphere increases substantially the list of companion candidates identified 
in similar surveys of B-type stars in the southern hemisphere. {\em Chandra} observations suggest that
most of such new, faint IR objects are truely bound companions. The detection of $\sim 60$\,\% of the B-type 
primaries with {\em Chandra} indicates the need to search for even closer, spectroscopic companions. Our pilot study 
has shown that it is possible to identify late-type stars in the IR spectrum of a B-type star.
For X-ray emitting B-type stars with negative results in all searches for companions, as an ultimate step 
sensitive X-ray spectroscopy shall be used to examine the physical conditions in the hot plasma. A comparison 
of these spectra with similar data for known late-type coronal X-ray sources will constrain the production 
mechanism for X-rays in intermediate-mass stars.  

The age dependence of dynamo action in intermediate-mass stars is studied by a comparison of
X-ray properties of the primaries in our sample of MS stars to the X-ray properties of HAeBe stars. 
We find that the detection fraction with {\em Chandra} is even higher for the latter ones ($\sim 80$\,\%).
This may be due to either of two reasons: (i) a shut-off of the magnetic dynamo at a critical 
as yet undefined point in the life of a B-/A-type star, or (ii) unidentified binary companions of HAeBe stars.

\begin{acknowledgements}

This research has made use of the SIMBAD database, operated at CDS, Strasbourg, France,
and use of data products from the Two Micron All Sky Survey, which is a joint project of 
the University of Massachusetts and the Infrared Processing and Analysis Center/California 
Institute of Technology, funded by the National Aeronautics and Space Administration and 
the National Science Foundation.

\end{acknowledgements}

\end{document}